\documentclass[prb,twocolumn,showpacs]{revtex4}
\usepackage{amsmath}
\usepackage{dcolumn}
\usepackage{graphicx}
\usepackage{bm}

\begin{document}

\title{Kondo screening coexisting with ferromagnetic order as a possible
ground state for Kondo lattice systems}
\author{Guang-Bin Li and Guang-Ming Zhang}
\affiliation{Department of Physics, Tsinghua University, Beijing 100084, China}
\author{Lu Yu}
\affiliation{Institute of Physics, Chinese Academy of Sciences, Beijing 100190, China;\\
Institute of Theoretical Physics, Chinese Academy of Sciences, Beijing
100190, China}
\date{\today }

\begin{abstract}
We consider the competition between the Kondo screening effect and
ferromagnetic long-range order\ (FLRO) within a mean-field theory of the
Kondo lattice model for low conduction electron densities $n_{c}$. Depending
on the parameter values, several types of FLRO ground states are found. When
$n_{c}<0.16$, a polarized FLRO phase is dominant in the large Kondo coupling
limit. For $0.16<n_{c}<0.82$, a non-polarized FLRO phase appears in the weak
Kondo coupling region; while in the intermediate coupling region the ground
state corresponds to the polarized and non-polarized FLRO phases,
respectively, \textit{coexisting} with the Kondo screening. For a strong
Kondo coupling, the product of pure Kondo singlets is the ground state.
Moreover, we also find that a weak magnetic field makes the pure Kondo
singlet phase vanish, while the non-polarized FLRO state \textit{with} the
Kondo screening spans a large area in the phase diagram.
\end{abstract}

\maketitle

The Kondo lattice model is usually considered as a theoretical model for
heavy fermion materials.\cite{Tsunetsugu1997} For this model, an important
issue arises from the interplay between the Kondo screening and the magnetic
interactions among local moments mediated by the conduction electrons,
namely, the Ruderman-Kittel-Kasuya-Yosida (RKKY) exchange interaction. The
former effect favors the formation of Kondo singlet state in the strong
coupling limit, while the latter interactions tend to stabilize a
magnetically long-range ordered state in the weak coupling limit. In-between
these two distinct phases, there exists a quantum phase transition. The
nature of such a transition has been a long standing issue since it was
first suggested by Doniach.\cite{Doniach1977} In our earlier paper,\cite%
{Zhang2000a} we proposed an extended mean field theory for an anisotropic
Kondo lattice model with half-filled conduction electrons. Since the
magnetic interactions and the Kondo screening were treated on an equal
footing, we found that the disordered Kondo singlet state can coexist with
antiferromagnetic long-range order (AFLRO) in the intermediate coupling
regime, where the local moments are partially screened, resulting in a very
small staggered magnetizations. These results have been confirmed in the
whole AFLRO phase by quantum Monte Carlo calculations.\cite%
{Assaad2001,Ogata2007} Thus, the proposed extended mean-field theory can
provide quite reliable physical results in the intermediate and strong Kondo
coupling region.

So far most of the theoretical studies in this field focus on the
antiferromagnetic heavy fermion materials. However, more and more heavy
fermion metals with a ferromagnetic long-range order (FLRO) have been found
experimentally: early examples are CePt$_{x}$Si,\cite{lee-ku-shelton} CeRu$%
_{2}$Ge$_{2}$,\cite{Sullow-1999} CePt,\cite{Larrea} CeSi$_{1.81}$,\cite%
{Drotziger} CeAgSb$_{2}$,\cite{Sidorov} and URu$_{2-x}$Re$_{x}$Si$_{2}$ at $%
x>0.3$ (Ref.\cite{Bauer-2005}). Recently new ferromagnetic heavy fermion
materials CeRuPO (Ref.\cite{Krellner}) and UIr$_{2}$Zn$_{20}$ (Ref. \cite%
{Bauer-2006}) have been discovered. For some uranium compounds, the $S=1$
underscreened Kondo lattice model has been proposed to explain the
coexistence of ferromagnetism and Kondo effect.\cite{Perkins} To account for
the FLRO state in the Kondo lattice model, we can actually assume that the
density of conduction electrons per local moment $n_{c}$ is less than one,
and then the FLRO states are more favorable energetically than the AFLRO
state.\cite{Sigrist-1992,Li1996} The interplay between the Kondo screening
and ferromagnetic long-range correlations is thus the most important issue.
Another related essential problem is the nature of the Fermi surface,\cite%
{Si} whether it is large, encompassing both the local moments and conduction
electrons, or it is small, incorporating only the conduction electrons.

In this paper, motivated by the success of our previous investigation,\cite%
{Zhang2000a} we will discuss these issues within the framework of a
mean-field theory. By introducing the mean field order parameters, the FLRO
and local Kondo screening effect can be treated on an equal footing, and we
find several types of FLRO ground phases. In particular, for $%
0.16<n_{c}<0.82 $, the ground state is a non-polarized FLRO phase in the
weak coupling limit; while in the intermediate coupling regime the ground
state can be polarized and non-polarized FLRO phases \textit{coexisting}
with a partial Kondo screening, depending on the Kondo coupling strength. In
those FLRO phases \textit{without} Kondo screening, the Fermi surface is a
small one; while for those FLRO states with Kondo screening, the Fermi
surface is a large one. Moreover, we also find that a weak magnetic field
will make the \textit{pure} Kondo singlet disordered state vanish and the
non-polarized FLRO state \textit{with} Kondo screening spanning a large area
in the phase diagram.

The model Hamiltonian of the Kondo lattice systems is defined by%
\begin{equation}
\mathcal{H}=\sum_{\mathbf{k},{\sigma }}\epsilon _{\mathbf{k}}c_{\mathbf{k}%
\sigma }^{\dagger }c_{\mathbf{k}\sigma }+J\sum_{i}\mathbf{\sigma }_{i}\cdot
\mathbf{S}_{i}-\sum_{i}\mathbf{B}\cdot \left( \mathbf{\sigma }_{i}+\mathbf{S}%
_{i}\right) ,
\end{equation}%
where $\epsilon _{\mathbf{k}}$ is the dispersion of the conduction
electrons, $\mathbf{\sigma }_{i}=\frac{1}{2}\sum_{{\alpha }\beta }c_{i\alpha
}^{\dagger }\mathbf{\tau }_{\alpha \beta }c_{i\beta }$ is the spin density
operator of the conduction electrons, $\mathbf{\tau }$ is the Pauli matrix,
the Kondo coupling strength $J>0$, and a magnetic field couples equally to
the local moments and conduction electrons. When the localized spins are
represented by $\mathbf{S}_{i}=\frac{1}{2}\sum_{{\alpha }\beta }f_{i\alpha
}^{\dagger }\mathbf{\tau }_{\alpha \beta }f_{i\beta }$ in the pseudo-fermion
representation, the projection into the physical subspace has to be
implemented by a local constraint $\sum_{\sigma }f_{i\sigma }^{\dagger
}f_{i\sigma }=1$. It is straightforward to decompose the Kondo spin exchange
into longitudinal and transversal parts
\begin{equation*}
\mathbf{\sigma }_{i}\cdot \mathbf{S}_{i}=\sigma _{i}^{z}S_{i}^{z}-\frac{1}{4}%
[(c_{i\uparrow }^{\dagger }f_{i\uparrow }+f_{i\downarrow }^{\dagger
}c_{i\downarrow })^{2}+(c_{i\downarrow }^{\dagger }f_{i\downarrow
}+f_{i\uparrow }^{\dagger }c_{i\uparrow })^{2}],
\end{equation*}%
where the longitudinal part describes the polarization of the conduction
electrons, giving rise to the usual RKKY interaction between the local
moments; while the transverse part represents the spin-flip scattering of
the conduction by the local moments, yielding the local Kondo screening
effect.\cite{Lacroix1979} The competition between these two interaction
parts determines the possible ground state of the Kondo lattice systems. To
correctly describe the ground state properties, both the Kondo screening
effect and RKKY magnetic correlation should be treated on an equal footing.

To develop a simple mean field theory, we introduce FLRO order parameters: $%
m_{f}=\left\langle S_{i}^{z}\right\rangle $ and $m_{c}=\langle \sigma
_{i}^{z}\rangle $ to decouple the longitudinal exchange term. Similarly, to
describe the Kondo screening effect, a hybridization order parameter $%
V=\langle c_{i\uparrow }^{\dagger }f_{i\uparrow }+f_{i\downarrow }^{\dagger
}c_{i\downarrow }\rangle $ is used to decouple the transverse exchange term.
We also introduce a Lagrangian multiplier $\lambda $ to enforce the local
constraint, which becomes the chemical potential in the mean field
approximation. Then the mean field Hamiltonian in the momentum space can be
written in a matrix form,%
\begin{equation}
\mathcal{H}_{MF}=\sum_{\mathbf{k},{\sigma }}\left( c_{\mathbf{k}\sigma
}^{\dagger },f_{\mathbf{k}\sigma }^{\dagger }\right) \left(
\begin{array}{cc}
\epsilon _{\mathbf{k}{\sigma }} & -\frac{JV}{2} \\
-\frac{JV}{2} & \lambda _{{\sigma }}%
\end{array}%
\right) \left(
\begin{array}{c}
c_{\mathbf{k}\sigma } \\
f_{\mathbf{k}\sigma }%
\end{array}%
\right) +\mathcal{N}\varepsilon _{0},
\end{equation}%
where $\epsilon _{\mathbf{k}{\sigma }}=\epsilon _{\mathbf{k}}+\frac{Jm_{f}-B%
}{2}{\sigma }$, $\lambda _{{\sigma }}=\lambda +\frac{Jm_{c}-B}{2}{\sigma }$,
$\varepsilon _{0}=\frac{JV^{2}}{2}-Jm_{c}m_{f}-\lambda $, ${\sigma =\pm 1}$
denote the up and down spin orientations, and $\mathcal{N}$ is the total
number of lattice sites. The quasiparticle excitation spectra are thus
obtained by
\begin{equation}
E_{\mathbf{k}{\sigma }}^{\pm }=\frac{1}{2}\left[ \epsilon _{\mathbf{k}{%
\sigma }}+\lambda _{{\sigma }}\pm \sqrt{\left( \epsilon _{\mathbf{k}{\sigma }%
}-\lambda _{{\sigma }}\right) ^{2}+(JV)^{2}}\right] ,
\end{equation}%
where there appear four quasiparticle bands with spin splitting.

By using the method of equation of motion, we derive the following single
particle Green functions
\begin{gather}
\langle \langle c_{\mathbf{k}\sigma }|c_{\mathbf{k}\sigma }^{\dagger
}\rangle \rangle _{\omega }=\frac{\omega -\lambda _{{\sigma }}}{(\omega
-\epsilon _{\mathbf{k}{\sigma }})(\omega -\lambda _{{\sigma }})-(\frac{JV}{2}%
)^{2}},  \notag \\
\langle \langle f_{\mathbf{k}\sigma }|f_{\mathbf{k}\sigma }^{\dagger
}\rangle \rangle _{\omega }=\frac{\omega -\epsilon _{\mathbf{k}{\sigma }}}{%
(\omega -\epsilon _{\mathbf{k}{\sigma }})(\omega -\lambda _{{\sigma }})-(%
\frac{JV}{2})^{2}},  \notag \\
\langle \langle f_{\mathbf{k}\sigma }|c_{\mathbf{k}\sigma }^{\dagger
}\rangle \rangle _{\omega }=\langle \langle c_{\mathbf{k}\sigma }|f_{\mathbf{%
k}\sigma }^{\dagger }\rangle \rangle _{\omega }=\frac{-JV/2}{\omega -\lambda
_{{\sigma }}}\langle \langle c_{\mathbf{k}\sigma }|c_{\mathbf{k}\sigma
}^{\dagger }\rangle \rangle .
\end{gather}%
Accordingly, the corresponding density of states can be calculated and
expressed as
\begin{gather}
\rho _{c}^{\sigma }(\omega )=\rho _{c}^{0}[\theta (\omega -\omega _{1\sigma
})\theta (\omega _{2\sigma }-\omega )+\theta (\omega -\omega _{3\sigma
})\theta (\omega _{4\sigma }-\omega )],  \notag \\
\rho _{f}^{\sigma }(\omega )=\left( \frac{JV/2}{\omega -\lambda _{{\sigma }}}%
\right) ^{2}\rho _{c}^{\sigma }(\omega ),  \notag \\
\rho _{f,c}^{\sigma }(\omega )=-\frac{JV/2}{\omega -\lambda _{{\sigma }}}%
\rho _{c}^{\sigma }(\omega ),
\end{gather}%
where $\theta (\omega )$ is a step function. When the density of states of
conduction electrons is assumed to be a constant $\rho _{c}^{0}=\frac{1}{2D}$
with $D$ as a half-width of the conduction electron band, the four
quasiparticle band edges can be expressed as
\begin{align*}
\omega _{1\sigma }& =\frac{1}{2}\left[ \epsilon _{\sigma }-D+\lambda _{{%
\sigma }}-\sqrt{(\epsilon _{\sigma }-D-\lambda _{{\sigma }})^{2}+(JV)^{2}}%
\right] , \\
\omega _{2\sigma }& =\frac{1}{2}\left[ \epsilon _{\sigma }+D+\lambda _{{%
\sigma }}-\sqrt{(\epsilon _{\sigma }+D-\lambda _{{\sigma }})^{2}+(JV)^{2}}%
\right] , \\
\omega _{3\sigma }& =\frac{1}{2}\left[ \epsilon _{\sigma }-D+\lambda _{{%
\sigma }}+\sqrt{(\epsilon _{\sigma }-D-\lambda _{{\sigma }})^{2}+(JV)^{2}}%
\right] , \\
\omega _{4\sigma }& =\frac{1}{2}\left[ \epsilon _{\sigma }+D+\lambda _{{%
\sigma }}+\sqrt{(\epsilon _{\sigma }+D-\lambda _{{\sigma }})^{2}+(JV)^{2}}%
\right] ,
\end{align*}%
where $\epsilon _{\sigma }=\frac{Jm_{f}-B}{2}{\sigma }$ and $\omega
_{1\sigma }<\omega _{2\sigma }<\omega _{3\sigma }<\omega _{4\sigma }$.
\begin{figure}[tbp]
\includegraphics[width=1.8in,angle=-90]{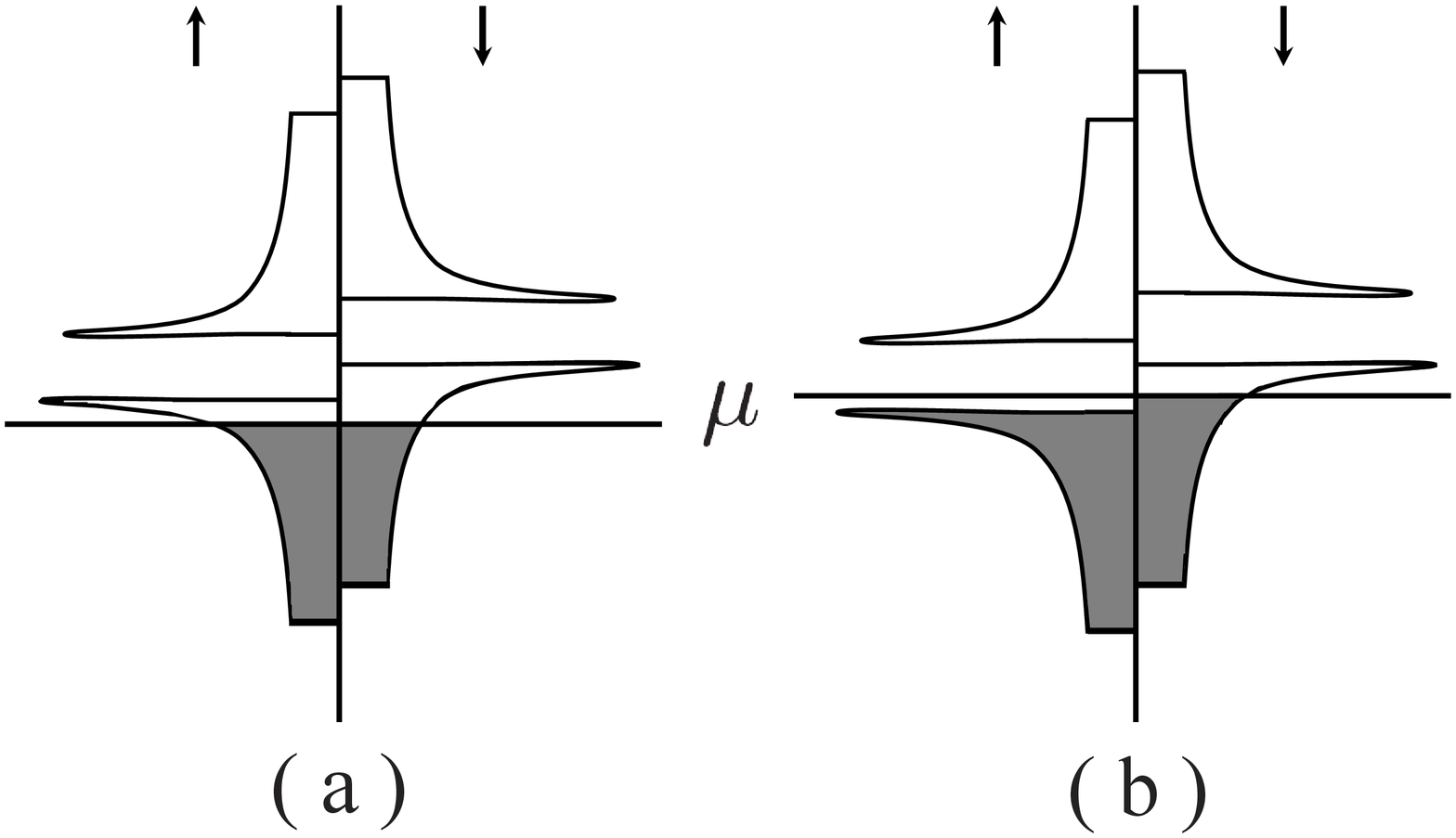}
\caption{Schematic plot of total DOS in the presence of Kondo screening
effect. (a) for the non-polarized FLRO phase, (b) for the polarized FLRO
phase.}
\label{fig-DOS}
\end{figure}

In the ground state, to keep the average numbers of $f$-electrons and
conduction electrons equal to $1$ and $n_{c}$, respectively, the chemical
potential $\mu $ has to be determined self-consistently. Therefore, the
self-consistent equations determining the various mean-field parameters $%
m_{c}$, $m_{f}$, $V$, $\lambda $, and the chemical potential $\mu $ are
given as follows
\begin{equation}
\left\{
\begin{array}{c}
n_{c} \\
2m_{c} \\
2V \\
1 \\
2m_{f}%
\end{array}%
\right\} =\sum_{\mathbf{\sigma }}\int_{-\infty }^{+\infty }d\omega \rho
_{c}^{\sigma }(\omega )\left\{
\begin{array}{c}
1 \\
\sigma \\
-\frac{JV}{\omega -\lambda _{{\sigma }}} \\
\left( \frac{JV/2}{\omega -\lambda _{{\sigma }}}\right) ^{2} \\
\sigma \left( \frac{JV/2}{\omega -\lambda _{{\sigma }}}\right) ^{2}%
\end{array}%
\right\} .
\end{equation}%
Here the position of the chemical potential $\mu $ with respect to the band
edges is an important factor. For $n_{c}<1$, there are two possible cases: $%
\omega _{1\downarrow }<\mu <\omega _{2\uparrow }$ and $\omega _{2\uparrow
}<\mu <\omega _{2\downarrow }$. Moreover, we should assume that $n_{c}<0.82$
so that the RKKY interaction between the nearest neighboring local moments
favors ferromagnetic coupling. \cite{Fazekas1991} Actually, the
corresponding results for $1<n_{c}<2$ can be obtained via the particle-hole
transformation $n_{c}\rightarrow \left( 2-n_{c}\right) $ and $\sigma
\rightarrow -\sigma $.

For $\omega _{1\downarrow }<\mu <\omega _{2\uparrow }$, the corresponding
density of states is schematically shown in Fig. \ref{fig-DOS}(a), where
both the lower spin-up and spin-down quasiparticle bands are partially
occupied. When we introduce the new variables,%
\begin{equation}
x_{\sigma }=\lambda _{\sigma }-\mu \text{, }y_{\sigma }=\lambda _{\sigma
}-\omega _{1\sigma },
\end{equation}%
the solution to the coupled self-consistent equations can be expressed in
terms of $m_{c}$ and $m_{f}$%
\begin{align}
x_{\uparrow }& =\frac{D(n_{c}+2m_{c})}{2\sinh (\frac{2D}{J})}\left[
e^{-2D/J}+\sqrt{\frac{(n_{c}-2m_{c})(1-2m_{f})}{(n_{c}+2m_{c})(1+2m_{f})}}%
\right] ,  \notag \\
y_{\uparrow }& =\frac{D(n_{c}+2m_{c})}{2\sinh (\frac{2D}{J})}\left[ e^{2D/J}+%
\sqrt{\frac{(n_{c}-2m_{c})(1-2m_{f})}{(n_{c}+2m_{c})(1+2m_{f})}}\right] ,
\notag \\
x_{\downarrow }& =\frac{D(n_{c}-2m_{c})}{2\sinh (\frac{2D}{J})}\left[
e^{-2D/J}+\sqrt{\frac{(n_{c}+2m_{c})(1+2m_{f})}{(n_{c}-2m_{c})(1-2m_{f})}}%
\right] ,  \notag \\
y_{\downarrow }& =\frac{D(n_{c}-2m_{c})}{2\sinh (\frac{2D}{J})}\left[
e^{2D/J}+\sqrt{\frac{(n_{c}+2m_{c})(1+2m_{f})}{(n_{c}-2m_{c})(1-2m_{f})}}%
\right] .
\end{align}%
Considering the relation $x_{\downarrow }-x_{\uparrow }=B-Jm_{c}$ and the
expressions for the quasiparticle band edges, we found a self-consistent
equation for $m_{c}$,%
\begin{equation}
\sqrt{n_{c}^{2}-4m_{c}^{2}+4(\eta -\alpha m_{c})^{2}}=\frac{\alpha (\alpha
m_{c}-\eta )-m_{c}}{2m_{c}\sinh (2D/J)-\eta },  \label{eq-mag-mc}
\end{equation}%
with $\alpha =\frac{J}{2D}\sinh \left( \frac{2D}{J}\right) -e^{-2D/J}$ and $%
\eta =\frac{B}{2D}\sinh (\frac{2D}{J})$. Solving Eq.(\ref{eq-mag-mc})
numerically will determine the magnetic order parameters $m_{c}$. Then $%
m_{f} $ can be calculated via%
\begin{equation}
m_{f}=\frac{\eta -2m_{c}\sinh (2D/J)}{\alpha +m_{c}/(\eta -\alpha m_{c})},
\end{equation}%
and the other mean-field parameters can also be obtained from their
equations as well.

However, when $\mathbf{B}=0$, we have $\eta =0$, and then analytical
solutions to $m_{c}$ and $m_{f}$ can be derived as,%
\begin{equation}
m_{c}=-\frac{1}{2}\sqrt{\frac{n_{c}^{2}-(\alpha \beta )^{2}}{1-\alpha ^{2}}}%
,m_{f}=-\frac{m_{c}}{\beta }
\end{equation}%
where $\beta =\frac{1-\alpha ^{2}}{2\alpha \sinh (\frac{2D}{J})}$ and the
hybridization parameter can be obtained
\begin{equation}
V^{2}=\frac{\cosh (\frac{2D}{J})\sqrt{\left( n_{c}^{2}-4m_{c}^{2}\right)
(\beta ^{2}-4m_{c}^{2})}+\beta n_{c}-4m_{c}^{2}}{2\beta \left( \frac{J}{2D}%
\sinh \frac{2D}{J}\right) ^{2}}.  \label{eq-V1}
\end{equation}

For $\omega _{2\uparrow }<\mu <\omega _{2\downarrow }$, the corresponding
density of states is schematically shown in Fig. \ref{fig-DOS}(b). Since the
lower spin up quasiparticle band is completely occupied, we have $%
n_{c\uparrow }+n_{f\uparrow }=1$. But the lower spin down quasiparticle band
is only partially occupied, the quasiparticles near the Fermi surface thus
becoming polarized with a total magnetization%
\begin{equation}
m_{c}+m_{f}=\frac{1-n_{c}}{2},
\end{equation}%
which corresponds to a plateau in the magnetization curve. Similarly, in
terms of the new variables: $x_{\uparrow }=\lambda _{\uparrow }-\omega
_{2\uparrow }$, $x_{\downarrow }=\lambda _{\downarrow }-\mu $, and $%
y_{\sigma }=\lambda _{\sigma }-\omega _{1\sigma }$, the solution to the
coupled self-consistent equations can be derived in terms of $m_{c}$ and $%
m_{f}$%
\begin{align}
x_{\uparrow }& =\frac{D(n_{c}+2m_{c})}{2\sinh (\frac{2D}{J})}\left[
e^{-2D/J}+\sqrt{\frac{n_{c}-2m_{c}}{1+2m_{f}}}\right] ,  \notag \\
y_{\uparrow }& =\frac{D(n_{c}+2m_{c})}{2\sinh (\frac{2D}{J})}\left[ e^{2D/J}+%
\sqrt{\frac{n_{c}-2m_{c}}{1+2m_{f}}}\right] ,  \notag \\
x_{\downarrow }& =\frac{D(n_{c}-2m_{c})}{2\sinh (\frac{2D}{J})}\left[
e^{-2D/J}+\sqrt{\frac{1+2m_{f}}{n_{c}-2m_{c}}}\right] ,  \notag \\
y_{\downarrow }& =\frac{D(n_{c}-2m_{c})}{2\sinh (\frac{2D}{J})}\left[
e^{2D/J}+\sqrt{\frac{1+2m_{f}}{n_{c}-2m_{c}}}\right] .
\end{align}%
When the relations $y_{\uparrow }-x_{\uparrow }=D(n_{c}+2m_{c})$, $%
y_{\downarrow }-x_{\downarrow }=D(n_{c}-2m_{c})$, $m_{c}+m_{f}=\frac{1-n_{c}%
}{2}$, and the band edge expressions are taken into account, the final forms
of the mean field order parameters can be obtained as%
\begin{gather}
m_{c}=-\frac{\gamma \left( 1-n_{c}\right) }{2\left( 1-\gamma \right) }%
,m_{f}=-\frac{m_{c}}{\gamma },  \notag \\
V^{2}=\frac{\cosh (\frac{2D}{J})\sqrt{\left( n_{c}-2m_{c}\right) (1+2m_{f})}%
+1-2m_{c}}{2(1-2m_{f})^{-1}\left( \frac{J}{2D}\sinh \frac{2D}{J}\right) ^{2}}%
,  \label{eq-V2}
\end{gather}%
where $\gamma $ has to be determined from the following equation
self-consistently
\begin{align}
& \sqrt{\frac{n_{c}\left( 1-2\gamma \right) +\gamma }{2-n_{c}-\gamma }}%
+\gamma \sqrt{\frac{2-n_{c}-\gamma }{n_{c}\left( 1-2\gamma \right) +\gamma }}
\notag \\
& =\alpha \left( 1+\gamma \right) -2\gamma \sinh \left( \frac{2D}{J}\right) .
\label{eq-gamma}
\end{align}%
Notice that this solution is \textit{independent} of the external magnetic
field and a constraint $\gamma <n_{c}$ has to be taken into account when
solving Eq.(\ref{eq-gamma}) numerically. \

Actually, it is easy to verify that the pure saturated FLRO state ($%
m_{f}=1/2 $) without a hybridization can also be a solution to Eq. (\ref%
{eq-V1}) and Eq. (\ref{eq-V2}). However, our detailed calculations lead to
the polarized and non-polarized FLRO phases, and we will call them polar-I
and non-polar-I FLRO states, respectively, where the conduction electrons
are partially or completely polarized by the saturated local moments with
magnetizations
\begin{equation}
\text{ }m_{c}=\left\{
\begin{array}{c}
-\frac{n_{c}}{2},\text{ \ \ \ \ }n_{c}<\frac{J-2B}{4D} \\
-\frac{J-2B}{8D},\text{ \ \ }n_{c}>\frac{J-2B}{4D}%
\end{array}%
\right. .
\end{equation}%
Obviously, the phase boundary between these two pure FLRO phases is given by
$n_{c}=\frac{J-2B}{4D}$.

Similarly, it can also be noticed that the pure Kondo singlet phase ($%
m_{c}=m_{f}=0$ but $V\neq 0$) as a solution to the self-consistent equations
requires $B=0$, namely, the pure Kondo singlet disordered state \textit{only}
exists in the absence of external magnetic field. The hybridization
parameter of the Kondo screening is given by
\begin{equation}
V=\sqrt{n_{c}}\left( \frac{J}{D}\sinh \frac{D}{J}\right) ^{-1}.
\end{equation}%
The boundary between the pure FLRO phase and FLRO coexisting with the Kondo
screening effect should be described by the condition $V=0$.

\begin{figure}[tbp]
\includegraphics[width=3.4in]{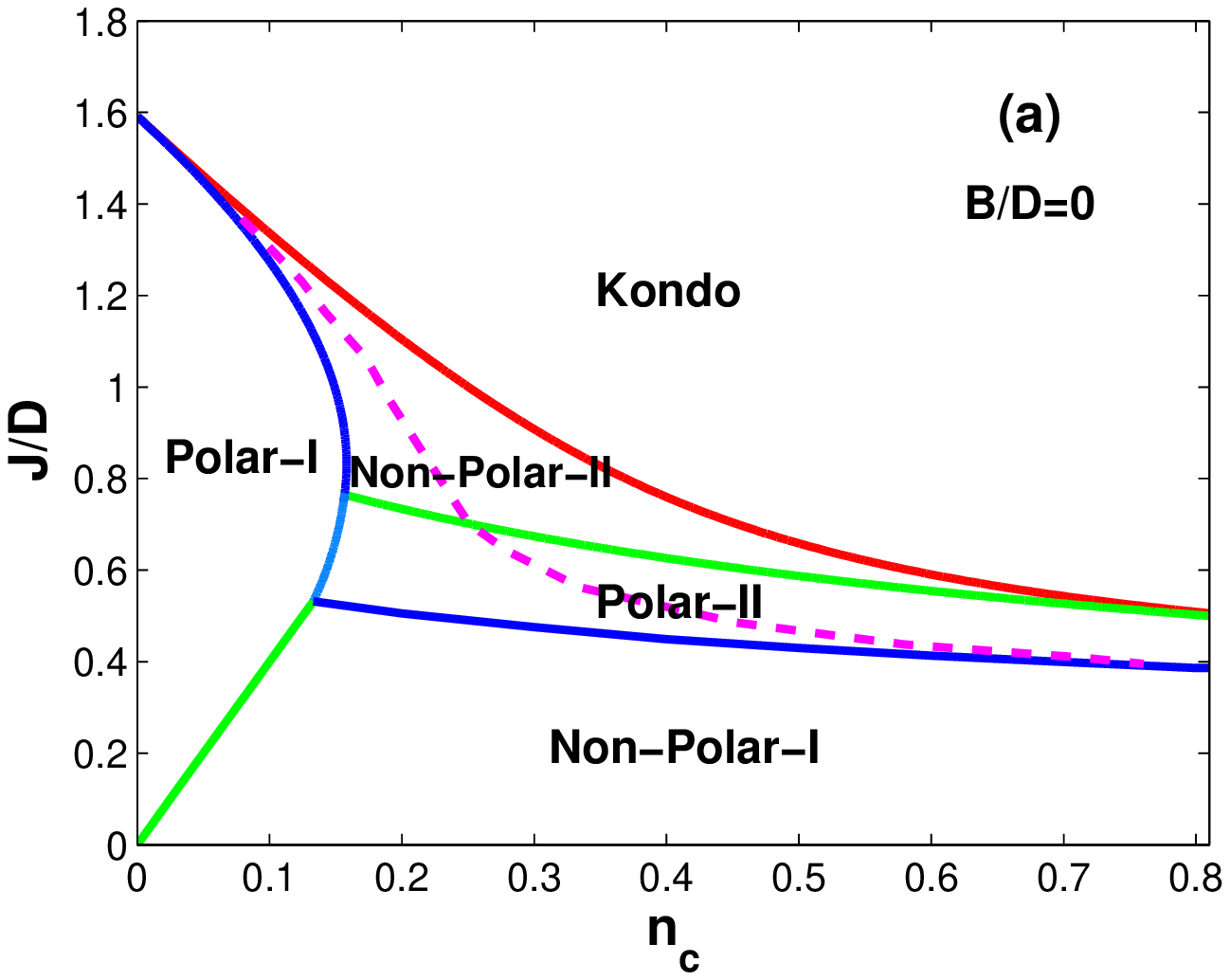}
\includegraphics
[width=3.4in]{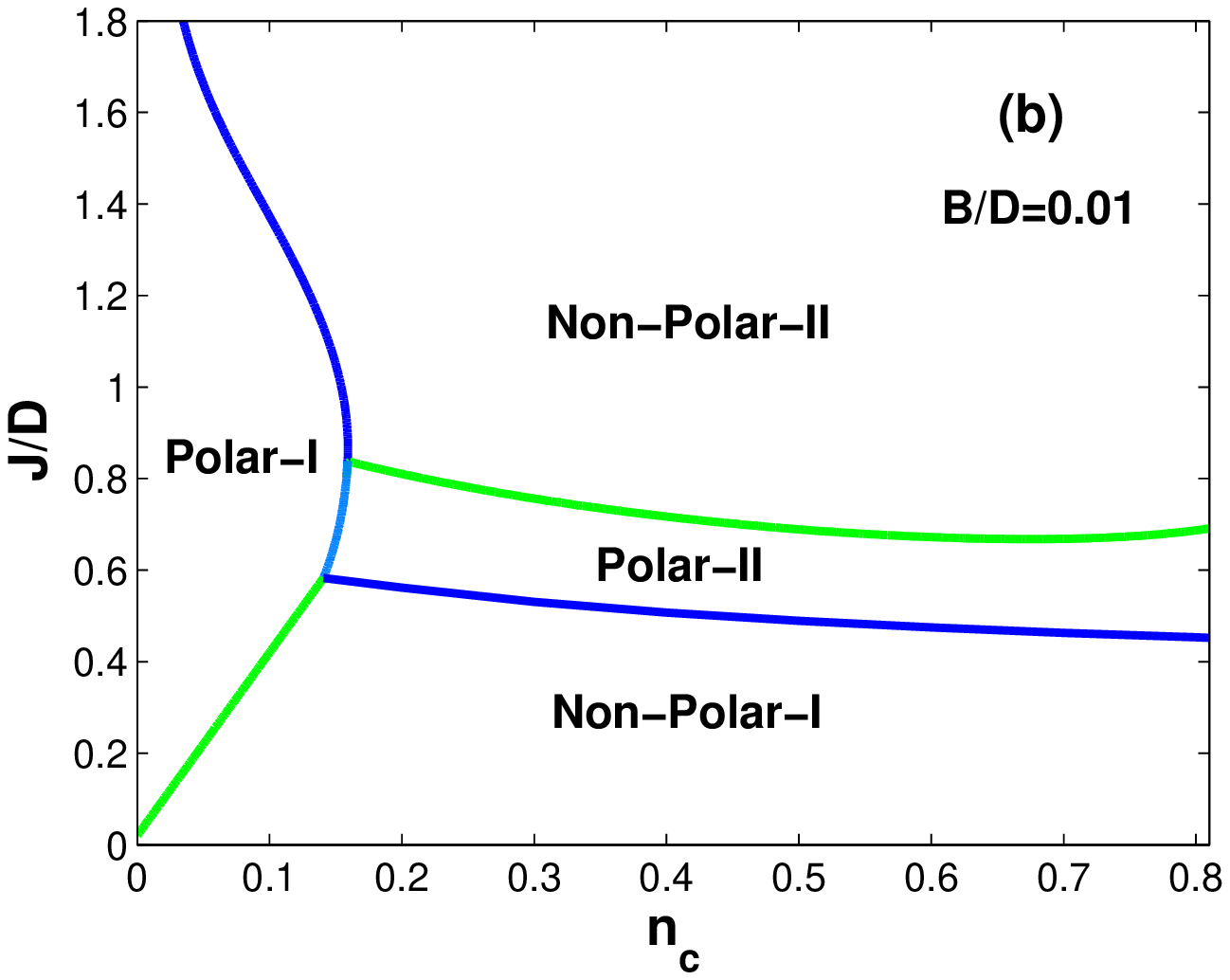}
\caption{(Color online) Ground state phase diagram of the Kondo lattice
model. (a) for B/D=0. The dashed line is given by comparing the ground state
energies of the pure Kondo singlet and pure FLRO states. (b) for B/D=0.01.}
\label{fig-pd}
\end{figure}

In general, we have to employ the numerical calculations to solve Eq.(\ref%
{eq-mag-mc}) and Eq.(\ref{eq-gamma}), respectively. The ground-state phase
diagram for $B=0$ is displayed in Fig. \ref{fig-pd}(a). Several types of
FLRO ground states can be found. When $n_{c}<0.16$, the polar-I FLRO phase
is dominant as a ground state in the large Kondo coupling region. For $%
0.16<n_{c}<0.82$, the ground state is given by the non-polar-I FLRO phase in
the weak Kondo coupling limit; while in the intermediate Kondo coupling
regime polarized and non-polarized FLRO phases with a finite value of the
hybridization parameter $V$ appear separately, depending on the Kondo
coupling strength. We will call them polar-II and non-polar-II FLRO states,
respectively. For a strong Kondo coupling, the pure Kondo singlet phase is
the ground state. The phase boundary between the polar-II and non-polar-II
FLRO phases is given by $2(m_{c}+m_{f})=1-n_{c}$, and the hybridization
parameter $V=0$ can be used to determine the phase boundaries between the
polar-I and polar-II FLRO phases as well as for the polar-I and non-polar-II
FLRO phases.

\begin{figure}[tbp]
\centering\includegraphics[width=3.4in]{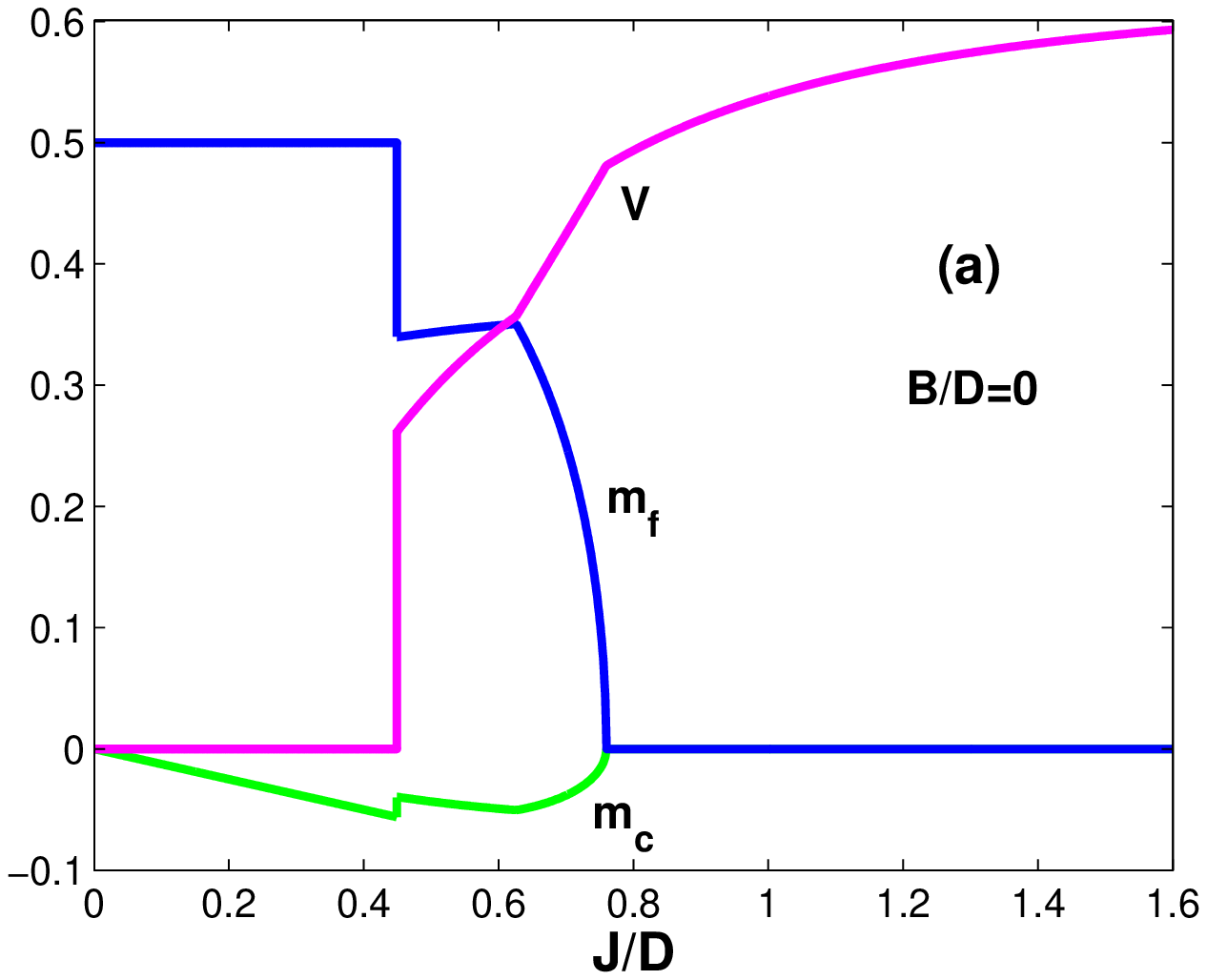}
\includegraphics
[width=3.4in]{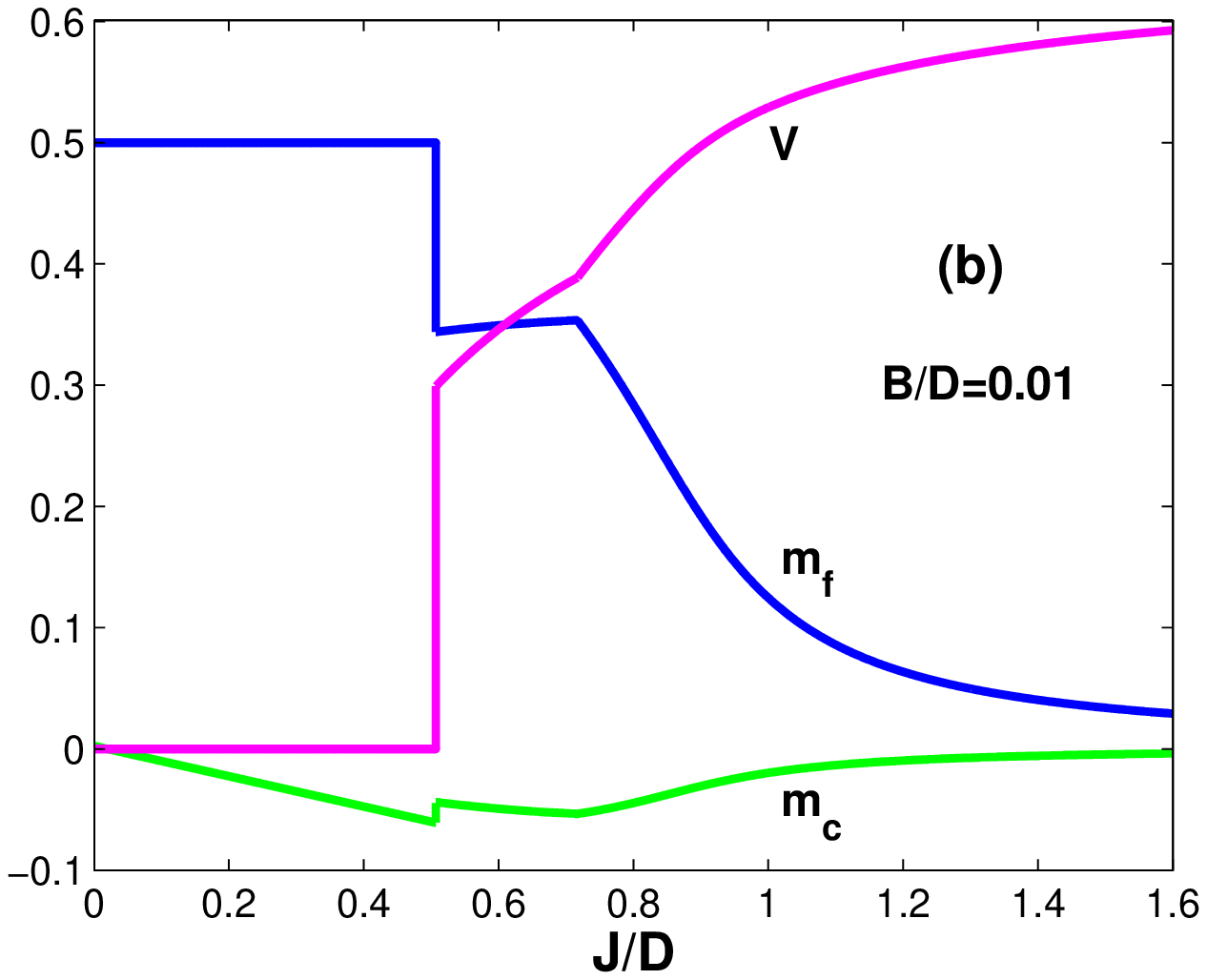}
\caption{(color online) Mean field order parameters as functions of $J/D$
for $n_{c}=0.4$. (a) $B/D=0$, (b) $B/D=0.01$.}
\label{fig-OP}
\end{figure}

\begin{figure}[tbp]
\includegraphics[width=3.40in]{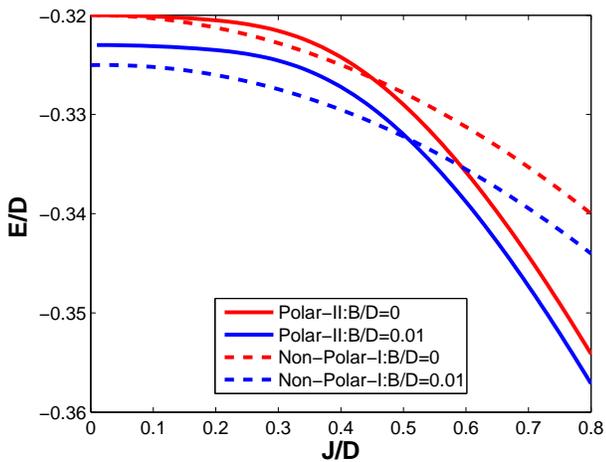}
\caption{(color online) Comparison of the ground state energies of
non-polar-I with polar-II FLRO phases in the absence and presence of a weak
magnetic field for $n_{c}=0.4$.}
\label{fig-GSE}
\end{figure}

For a given value of $n_{c}$, the mean field order parameters $m_{c}$, $%
m_{f} $, and $V$ as functions of the Kondo coupling strength are calculated
and shown in Fig. \ref{fig-OP}(a). All the quantum phase transitions in the
phase diagram are continuous second order, except for the phase transition
from non-polar-I to the polar-II FLRO phases, because the hybridization
parameter $V$ has a jump from zero to a finite value at the transition line.
The phase boundary of this first-order transition is actually determined by
comparing the two ground state energies, displayed in Fig. \ref{fig-GSE}.

\

\begin{figure}[tbp]
\includegraphics [width=3.4in]{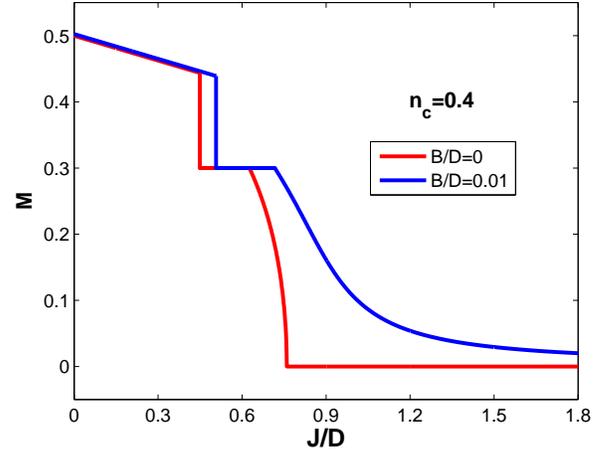}
\caption{(color online) Total magnetization curve as a function of $J/D$ for
a given value of $n_{c}$. A plateau appears when polarized quasiparticles
are present near Fermi surface.}
\label{fig-Mag2}
\end{figure}

The total magnetization $M$ as a function of the Kondo coupling strength for
$n_{c}=0.4$ is shown in Fig.\ref{fig-Mag2}, where a plateau appears in the
curve with a fixed value of $(1-n_{c})/2$. The onset position of this
plateau is just linked to the vanishing of the Fermi surface in the lower
spin up quasiparticles.\cite{Beach2008} Thus, the presence of such a plateau
can be used as an indication of the polarized quasiparticles near the Fermi
surface.

In the absence of the magnetic field, a very appealing physical picture
emerges. In the pure polarized and non-polarized FLRO phases, the conduction
electrons are decoupled from the local moments. For $0.16<n_{c}<0.82$, when $%
J/D$ is large enough, the localized moments are screened by the
conduction electrons via the formation of local singlets, leading
to a product of the local Kondo spin singlet disordered phase. It
should be noticed that those conduction electrons in the Kondo
singlets are \textit{not localized but itinerant} so that the
Kondo lattice system still displays metallic properties. As $J/D$
is in the intermediate coupling regime, the local moments are only
partially screened by the conduction electrons, and the remaining
uncompensated parts on the neighboring lattice sites develop the
ferromagnetic long-range correlations mediated by the conduction
electron spins, yielding the FLRO phases. The resulting FLRO
states are either polarized or non-polarized, depending on the
Kondo coupling strength. In these FLRO phases \textit{with} Kondo
screening, the quasiparticles consist of the conduction electrons
and local moments, and the corresponding Fermi surface should be
the large one. However, as $J/D$ becomes much smaller in the same
density range, the local moments decoupled from the conduction
electrons are aligned, forming the long-range order. In those FLRO
phases \textit{without} Kondo screening, the Fermi surface only
involves the conduction electrons, corresponding to a small Fermi
surface. Therefore, the hybridization between the conduction
electrons and the $f$-electrons $V\neq 0$ or $V=0$ will determine
whether the Fermi surface is large or small.

In the presence of a weak magnetic field, the mean field order parameters $%
m_{c}$, $m_{f}$, and $V$ as functions of the Kondo coupling strength are
also calculated and shown in Fig. \ref{fig-OP}(b), where the pure Kondo
singlet disordered phase vanishes completely. The corresponding phase
diagram is displayed in Fig. \ref{fig-pd}(b). Generally speaking, in
response to a weak magnetic field, the Zeeman splitting lifts the degeneracy
of the spin up and spin down in the hybridization channel, shifting them
with respect to one another and producing the magnetic moments. However,
both polar-II and non-polar-II FLRO phases still have a small but finite
value of $V$. Thus, the Kondo screening effect coexist with the FLRO phases
in the large Kondo coupling regime. In contrary, it has been demonstrated
that there is a critical magnetic field under which a Kondo singlet state
persists in the Kondo insulating ground state with AFLRO correlations. \cite%
{Beach2004,Ohashi2004}

In conclusion, due to the competition between the Kondo screening and
magnetic RKKY interactions in the low density of the conduction electrons, a
polar- and non-polar FLRO states can coexist with the Kondo screening both
in the absence and presence of a weak magnetic field. These two phases
should have large Fermi surfaces. On the other hand, the pure polar- and
non-polar FLRO phases have small Fermi surfaces. The applied weak magnetic
field makes the pure disordered Kondo singlet phase vanish. Moreover, there
also exist two polarized FLRO phases with the total magnetization fixed by a
value $M=(1-n_{c})/2$. To some extent, our present mean field theory
captures the physics of the Kondo lattice systems with ferromagnetic
correlations, especially the essence of the competition between Kondo
screening effect and ferromagnetism. In order to put these mean field
results on a solid ground, further investigations beyond the mean field
theory are certainly needed.

This work is partially supported by NSF-China and the National Program for
Basic Research of MOST, China.


\begin{thebibliography}{99}
\bibitem{Tsunetsugu1997} H. Tsunetsugu, M. Sigrist, and K. Ueda, Rev. Mod.
Phys. \textbf{69}, 809 (1997), and references therein.

\bibitem{Doniach1977} S. Doniach, Physica, B \& C \textbf{91}, 231 (1977).

\bibitem{Zhang2000a} G. M. Zhang and L. Yu, Phys. Rev. B \textbf{62}, 76
(2000).

\bibitem{Assaad2001} S. Capponi and F. F. Assaad, Phys. Rev. B \textbf{63},
155114 (2001).

\bibitem{Ogata2007} H. Watanabe and M. Ogata, Phys. Rev. Lett. \textbf{99},
136401 (2007).

\bibitem{lee-ku-shelton} W. H. Lee, H. C. Ku, and R. N. Shelton, Phys. Rev.
B \textbf{38}, 11562 (1988).

\bibitem{Sullow-1999} S. Sullow, M. C. Aronson, B. D. Rainford, and P. Haen,
Phys. Rev. Lett. \textbf{82}, 2963 (1999).

\bibitem{Larrea} J. Larrea, et. al., Phys. Rev. B \textbf{72}, 035129 (2005).

\bibitem{Drotziger} S. Drotziger, et. al., Phys. Rev. B \textbf{73}, 214413
(2006).

\bibitem{Sidorov} V. A. Sidorov, et. al., Phys. Rev. B \textbf{67}, 224419
(2003).

\bibitem{Bauer-2005} E. D. Bauer, et. al., Phys. Rev. Lett. \textbf{94},
046401 (2005).

\bibitem{Krellner} C. Krellner, et. al., Phys. Rev. B \textbf{76}, 104418
(2007).

\bibitem{Bauer-2006} E. D. Bauer, et. al., Phys. Rev. B \textbf{74}, 155118
(2006).

\bibitem{Perkins} N. B. Perkins, J. R. Iglesias, M. D. Nunez-Regueiro, B.
Coqblin, Eur. Phys. Lett. \textbf{79}, 57006 (2007).

\bibitem{Sigrist-1992} M. Sigrist, K. Ueda, and H. Tsunetsugu, Phys. Rev. B
\textbf{46}, 175 (1992).

\bibitem{Li1996} Z. Z. Li, M. Zhuang, and M. W. Xiao, J. Phys.:Condens.
Matter \textbf{8}, 7941 (1996).

\bibitem{Si} S. J. Yamamoto and Q. Si, arXiv:0812.0819.

\bibitem{Lacroix1979} C. Lacroix, and M. Cyrot, Phys. Rev. B \textbf{20},
1969 (1979).

\bibitem{Fazekas1991} P. Fazekas and E. M\"{u}ller-Hartmann, Z. Phys. B
\textbf{85}, 285 (1991).

\bibitem{Beach2008} K. S. D. Beach and F. F. Assaad, Phys. Rev. B \textbf{77}%
, 205123 (2008).

\bibitem{Beach2004} K. S. D. Beach, P. A. Lee, and P. Monthoux, Phys. Rev.
Lett. \textbf{92}, 26401 (2004).

\bibitem{Ohashi2004} T. Ohashi, A. Koga, S. Suga, and N. Kawakami, Phys.
Rev. B \textbf{70}, 245104 (2004).
\end{thebibliography}
\end{document}